\documentclass[twocolumn,showpacs,preprintnumbers,amsmath,amssymb]{revtex4}

\usepackage{color}
\usepackage{graphics,epsfig,graphicx}
\usepackage{pstcol,times,graphicx,graphics,pstricks,pst-node,pst-coil}

\usepackage{dcolumn}
\usepackage{bm}
\usepackage{amsmath,amssymb}

\def\be{\begin{equation}}
\def\ee{\end{equation} }
\def\beq{ \begin{eqnarray}}
\def\eeq{\end{eqnarray} }
\def\bc{ \begin{center}}
\def\ec{\end{center} }

\begin{document}

\title{Particle creation in a Robertson-Walker Universe revisited}

\author{F. Pascoal}
\email{fabiopr@if.ufrj.br}
\affiliation{Departamento de F\'isica, Universidade Federal de S\~ao Carlos\\
\small Via Washington Luis, km 235. S\~ao Carlos, 13565-905, SP,
Brazil}

\author{C. Farina}%
\email{farina@if.ufrj.br}
\affiliation{%
Instituto de Física, Universidade Federal do Rio de Janeiro\\
Caixa Postal 68528, 21941-972, Rio de Janeiro, RJ, Brazil }

\date{\today}

\begin{abstract}
We reanalyze the problem of particle creation in a 3+1 spatially
closed Robertson-Walker space-time. We compute the total number of
particles produced by this non-stationary gravitational background
as well as the corresponding total energy and find a slight
discrepancy between our results and those recently obtained in the
literature \cite{SetareRW}.
\end{abstract}
 \maketitle

\section{Introduction}

In a curved space-time the concept of particles is a subtle one and
the meaning of what is a particle or what is a detector becomes much more
difficult than in a flat space-time \cite{BirrelDavies} . Basically, this happens because
in a general curved space-time we do not have Lorentz symmetry
anymore and it is precisely this symmetry that, in a flat space-time,
allows us to identify the best vacuum state for the
theory. Lorentz symmetry assures that the
vacuum state is the same for all inertial observers.

Fortunately, in many problems of our interest, there are regions
where a choice of the physical vacuum is quite natural. That happens
when the time dependence of the field in these regions is harmonic
(or, at least, almost harmonic). A basic condition to have a field
with a harmonic time dependence is that the background metric and
the eventual boundaries do not depend on time. In some specific
cases, the background metric respects this condition only in
asymptotic times (distant future and remote past). In these cases,
the physical interpretation of particles is possible at those
asymptotical times, but not between them. As a consequence,  a
non-static curved space-time background may lead to the phenomenon
of particle creation.

The first one to discuss the problem of particle creation due to a
curved space-time background was Schr\"odinger in
1939\cite{Schrodinger39}, but the first one that carefully
investigated this phenomenon was Parker in the late 60´s
\cite{Parker}. The particle creation in a $1+1$ spatially closed
Robertson-Walker space-time was investigated in
\cite{CriacaoparticulasRW,BirrelDavies}. The $3+1$ version of this problem
has been investigated in a recent work \cite{SetareRW}
\footnote
{This problem also was investigated in
\cite{MostepanenkoTrunov97}, but the authors have not written a
closed analytical expression for the total number of particles
created.}.
However, when generalizing the 1+1 solution to the 3+1
one, this author missed a degeneracy factor leading to an
incomplete answer for the total number of particles created as well as the
corresponding total energy.

The main purpose of this short paper is to compute the total number
of particles created and the total energy related to them taking
into account the correct degeneracy factor that arises when we are
in a 3+1 dimensions. With the aid of appropriated graphics, we also
make a brief discussion on how the parameters that appear in the
metric affect the total number of particles produced.

\section{Particle creation}

We consider the case of an expanding spatially closed
Robertson-Walker universe whose line element and scalar curvature
are given, respectively, by
\beq ds^2&=&a^2(\eta)
\Bigl(d\eta^2-d\rho^2-\,\mbox{sen}^2\rho(d\theta^2+\,\mbox{sen}^2
\theta d\varphi)\Bigr),\cr\cr
 R&=&6a^{-3} \left(\partial^2_\eta a +a\right),
\eeq
where $a(\eta)$ is the scale factor, $\eta$ is the conformal time
and $0\le \rho \le \pi$. In this case the equation of a massive
scalar field conformally coupled to the metric is written as
\begin{widetext}
\beq \frac{\partial^2 u_{\bf k}}{\partial \eta^2}+\frac{2}{a}
\frac{\partial a}{\partial \eta} \frac{\partial u_{\bf k}}{\partial
\eta}+\left( m^2 a^2+1 +\frac{1}{a} \frac{\partial^2 a}{\partial
\eta^2}\right) u_{\bf k}-\nabla_{ang} u_{\bf k}=0\label{eqdifCERW},
\eeq
where $\nabla_{ang}$ is the angular part of the Laplacian operator
on a 3-sphere. This operator has the hiper-spherical harmonics as
eigenfunctions, that satisfy the following differential equation
\cite{HHesf}
\be
\nabla_{ang}
\mbox{Y}(l, m_1, m_2;\varphi, \theta,\rho) =-l(l+2) \mbox{Y}(l, m_1,
m_2;\varphi, \theta,\rho),\label{eqdhiperesfericos}
\ee
\end{widetext}
where $l=0,\,1,\,2,\,...$, $m_1=0,\,\pm 1,\,\pm 2,\,...,\,\pm l$ and
$m_2=0,\,\pm 1,\,\pm 2,\,...,\,\pm m_1$. Hence, the $\nabla_{ang}$
in equation (\ref{eqdifCERW}) suggests the {\it ansatz},
\be
u_{{\bf k}}(\varphi,\theta,\rho,\eta)=\frac{1}{a(\eta)}\mbox{Y}(l, m_1,
m_2;\varphi, \theta,\rho) g_l(\eta). \label{Ansatz} \ee Substituting
(\ref{Ansatz}) in (\ref{eqdifCERW}), we obtain the differential
equation for $g_l(\eta)$, \be\frac{d^2 g_l(\eta)}{d
\eta^2}+\omega_l^2(\eta)\,g_l(\eta)=0, \label{eqdifg}
\ee
where
 \be
\omega_l^2(\eta)=(l+1)^2+m^2a^2(\eta).
\ee
Note that this equation is similar to that of a mechanical harmonic
oscillator with a time-dependent frequency.

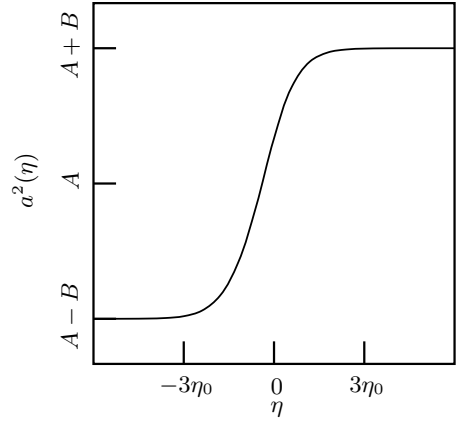
\begin{figure}[!h]
\begin{center}
\newpsobject{showgrid}{psgrid}{subgriddiv=1,griddots=10,gridlabels=6pt}
\begin{pspicture}(-3,-0.6)(2.4,4.8)
\psset{unit=0.6}
\pspolygon[linewidth=.05] (-4,0)(4,0)(4,8)(-4,8)
\pscurve[linewidth=.04] (-4., 1.000147449) (-3.600000000,
1.000489460) (-3.200000000, 1.001624061) (-2.800000000, 1.005381122)
(-2.400000000, 1.017747208) (-2., 1.057679493) (-1.600000000,
1.179747185) (-1.200000000, 1.508859069) (-.8000000000, 2.234343176)
(-.4000000000, 3.480058964) (0., 5.) (.4000000000, 6.155419532)
(.8000000000, 6.708778620) (1.200000000, 6.908208473) (1.600000000,
6.971957873) (2., 6.991517295) (2.400000000, 6.997441720)
(2.800000000, 6.999229158) (3.200000000, 6.999767799) (3.600000000,
6.999930060) (4., 6.999978934)
\psline[linewidth=.05](-4,7)(-3.5,7)
\psline[linewidth=.05](-4,4)(-3.5,4)
\psline[linewidth=.05](-4,1)(-3.5,1)
\psline[linewidth=.05](0,0)(0,0.5)
\psline[linewidth=.05](-2,0)(-2,0.5)
\psline[linewidth=.05](2,0)(2,0.5)
\rput{90}(-4.5,7){{ $A+B$}} \rput{90}(-4.5,1){{ $A-B$}}
\rput{90}(-4.5,4){{ $A$}} \rput{90}(-5.5,4){{ $a^2(\eta)$}}
\rput(0,-0.5){{ $0$}} \rput(-2,-0.5){{ $-3\eta_0$}} \rput(2,-0.5){{
$3\eta_0$}} \rput(0,-1){{ $\eta$}}
\end{pspicture}
\caption{Graphic of $a^2(\eta)$ {\it versus} $\eta$ for a closed
expanding universe.} \label{fig1}
\end{center}
\end{figure}

Now, we consider an exactly solvable case due to a convenient choice
of $a(\eta)$, namely
\be
a(\eta)=\sqrt{A+B\,\mbox{tanh}\left(\frac{\eta}{\eta_0}\right)},\label{fcRW}
\ee
where $A$, $B$ and $\eta_0$ are constants and $A>B$. Since we are
considering an expanding universe, $B
> 0$ and $\eta_0>0$. An inspection in Figure \ref{fig1} allow us to interpret $\eta_0$ as
the time scale of the expansion of the Universe while $A-B$ and
$A+B$ are, respectively, the scales of the size of the universe
before and after the expansion.

\begin{figure}[!h]
\newpsobject{showgrid}{psgrid}{subgriddiv=1,griddots=10,gridlabels=6pt}
\begin{center}
\begin{pspicture}(-0.6,-0.6)(4.8,4.8)
\psset{unit=0.8} \psset{arrowsize=0.1 1}
\pspolygon[linewidth=.05](0,0)(6,0)(6,6)(0,6)
\rput{90}(-0.5, 5.4604){ $N(m_0)$} \rput{90}(-0.5,0){ $0$}
\rput{90}(-1,3){ $N$}
\psline[linewidth=.05](0.375, 5.4604)(0, 5.4604)
\psline[linewidth=.05](1.2,0)(1.2,0.375) \rput(1.2,-0.5){ $m_0$}
\psline{<->}(0.5, 2.0087)(2.56,2.0087) \rput(1.53,2.4){ $\Delta
m_0$}
\rput(0,-0.5){ $0$} \rput(3,-1){ $m$}
\pscurve [linewidth=.04](0., 0.) (.05, .0029372) (.1000000000,
.034928) (.1500000000, .12503) (.2000000000, .28206) (.2500000000,
.50380) (.3000000000, .78324) (.3500000000, 1.1106) (.4000000000,
1.4746) (.4500000000, 1.8632) (.5000000000, 2.2646) (.5500000000,
2.6676) (.6000000000, 3.0624) (.6500000000, 3.4404) (.7000000000,
3.7946) (.7500000000, 4.1198) (.8000000000, 4.4118) (.8500000000,
4.6684) (.9000000000, 4.8880) (.9500000000, 5.0702) (1., 5.2156)
(1.050000000, 5.3254) (1.100000000, 5.4014) (1.150000000, 5.4456)
(1.200000000, 5.4604) (1.250000000, 5.4482) (1.300000000, 5.4120)
(1.350000000, 5.3540) (1.400000000, 5.2768) (1.450000000, 5.1832)
(1.500000000, 5.0754) (1.550000000, 4.9556) (1.600000000, 4.8258)
(1.650000000, 4.6882) (1.700000000, 4.5442) (1.750000000, 4.3956)
(1.800000000, 4.2436) (1.850000000, 4.0896) (1.900000000, 3.9348)
(1.950000000, 3.7800) (2., 3.6262) (2.050000000, 3.4740)
(2.100000000, 3.3242) (2.150000000, 3.1770) (2.200000000, 3.0332)
(2.250000000, 2.8928) (2.300000000, 2.7566) (2.350000000, 2.6242)
(2.400000000, 2.4962) (2.450000000, 2.3726) (2.500000000, 2.2534)
(2.550000000, 2.1388) (2.600000000, 2.0286) (2.650000000, 1.9230)
(2.700000000, 1.8218) (2.750000000, 1.7250) (2.800000000, 1.6324)
(2.850000000, 1.5441) (2.900000000, 1.4598) (2.950000000, 1.3796)
(3., 1.3031) (3.050000000, 1.2305) (3.100000000, 1.1614)
(3.150000000, 1.0958) (3.200000000, 1.0335) (3.250000000, .97448)
(3.300000000, .91852) (3.350000000, .86550) (3.400000000, .81528)
(3.450000000, .76778) (3.500000000, .72284) (3.550000000, .68034)
(3.600000000, .64020) (3.650000000, .60226) (3.700000000, .56644)
(3.750000000, .53264) (3.800000000, .50074) (3.850000000, .47064)
(3.900000000, .44228) (3.950000000, .41552) (4., .39032)
(4.050000000, .36658) (4.100000000, .34422) (4.150000000, .32316)
(4.200000000, .30334) (4.250000000, .28468) (4.300000000, .26714)
(4.350000000, .25064) (4.400000000, .23512) (4.450000000, .22052)
(4.500000000, .20680) (4.550000000, .19390) (4.600000000, .18179)
(4.650000000, .17041) (4.700000000, .15972) (4.750000000, .14968)
(4.800000000, .14025) (4.850000000, .13140) (4.900000000, .12310)
(4.950000000, .11530) (5., .10799) (5.050000000, .10113)
(5.100000000, .094690) (5.150000000, .088654) (5.200000000, .082994)
(5.250000000, .077686) (5.300000000, .07271) (5.350000000, .068046)
(5.400000000, .063676) (5.450000000, .059578) (5.500000000, .05574)
(5.550000000, .052144) (5.600000000, .048776) (5.650000000, .045620)
(5.700000000, .042666) (5.750000000, .039898) (5.8, .037308)
(5.850000000, .034882) (5.900000000, .032612) (5.950000000, .030486)
(6., .028496)
\end{pspicture}
\end{center}
\caption{Graphic of the total number of particles produced {\it
versus} the mass related to the field.} \label{fig2}
\end{figure}
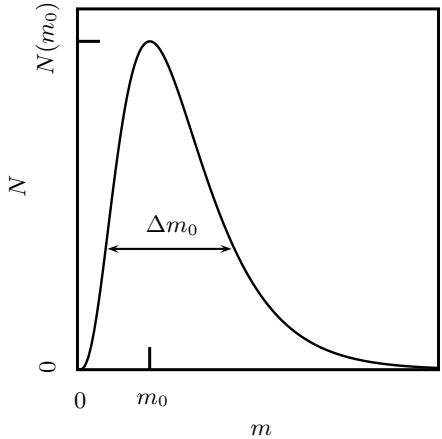

As we can see from Figure \ref{fig1}, there are two asymptotic times
where the background metric is almost static, namely, the remote
pass, characterized by $\eta \ll - \eta_0$ and the distant future,
characterized by $\eta \gg \eta_0$.  At those asymptotic times the
corresponding vacuum states may be well defined if the solutions of
(\ref{eqdifg}) have the following asymptotic limit
\begin{widetext}
\beq \lim_{\eta\to \mp\infty}
g_l^{(p,f)}(\eta)&=&\frac{e^{-i\omega_l^{(p,f)}\eta}}{\sqrt{2
\omega_l^{(p,f)}N_{\bf k}}}, \cr\cr
\omega_l^{(p,f)}&=&\sqrt{(l+1)^2+m^2(A\mp B)},\label{assimp} \eeq
where the superscripts $p$ and $f$ stand for past and future
solutions, respectively. The solutions of (\ref{eqdifg}) that
respect the asymptotic limits (\ref{assimp}) are given
,respectively, by \beq
g_l^{p}&=&\frac{\xi^{\frac{i}{2}\omega_l^{f}\eta_0}
(1-\xi)^{-\frac{i}{2}\omega_l^{p}\eta_0}}{\left(2\omega_l^{p} N_{\bf
k}\right)^{1/2}}\,_2F_1 \left(1+i\omega_l^{-} \eta_0,i\omega_l^{-}
\eta_0;1-i\omega_l^{p} \eta_0;1-\xi\right),\cr\cr
g_l^{f}&=&\frac{\xi^{\frac{i}{2}\omega_l^{f}\eta_0}
(1-\xi)^{-\frac{i}{2}\omega_l^{p}\eta_0}}{\left(2\omega_l^{f} N_{\bf
k}\right)^{1/2}}\,_2F_1 \left(1+i\omega_l^{-} \eta_0,i\omega_l^{-}
\eta_0;1+i\omega_l^{f} \eta_0;\xi\right), \eeq
 where $\,_2F_1$ is the hipergeometric function \cite{Hipergeometrica} and we have
defined  $\omega_l^{(\pm )}:=\frac{1}{2}\left( \omega_l^{f} \pm
\omega_l^{p} \right)$ and $
\xi:=\left(1-e^{2\eta/\eta_0}\right)^{-1}$. Since $u_{\bf k}^{p}$
and $u_{\bf k}^{f}$ are not equal, the corresponding Bogolubov
coefficients must be non-vanishing. Using some well known
proprieties of the hypergeometric function, we can write $u_{\bf
k}^{p}$ in terms of $u_{\bf k}^{f}$ and $u_{\bf k}^{f*}$ as follows
\be
u^{p}_{{\bf k}}= \sum_{{\bf k}'} \left( \alpha_{{\bf k} {\bf
k}'} u^{f}_{{\bf k}'} + \beta_{{\bf k} {\bf k}'} u^{f\,*}_{{\bf
k}'}\right), \ee with the Bogolugov coefficients given by\beq
\alpha_{\bf k' k }&=&\delta_{\bf k
k'}\sqrt{\frac{\omega_l^{f}}{\omega_l^{p}
}}\frac{\Gamma(1-i\omega_l^{p} \eta_0) \Gamma(-i\omega_l^{f}
\eta_0)}{\Gamma(-i\omega_l^{+} \eta_0) \Gamma(1-i\omega_l^{+}
\eta_0)};\cr\cr  \beta_{\bf k' k }&=&\delta_{\bf k
k'}\sqrt{\frac{\omega_l^{f}}{\omega_l^{p}
}}\frac{\Gamma(1-i\omega_l^{p} \eta_0) \Gamma(i\omega_l^{f}
\eta_0)}{\Gamma(i\omega_l^{-} \eta_0) \Gamma(1+i\omega_l^{-}
\eta_0)} . \label{BogolugovRW}\eeq

In the remote past all the inertial particle detectors register the
complete absence of particles in the state $\left| 0^p\right>$ (the
vacuum state associated to $u_{\bf k}^{p}$). However, in the distant
future, any inertial particle detector will register a number of
particles with quantum numbers $\bf k$ in the $\left| 0^p\right>$
state given by
\be
\left<0^p\right|N^f_{\bf
k}\left|0^p\right>=\sum_{\bf k'} \left| \beta_{\bf k
k'}\right|^2=\frac{\mbox{sinh}^2(\pi \omega_l^{(-)}
\eta_0)}{\mbox{sinh}(\pi \omega_l^{p} \eta_0) \,\mbox{sinh}(\pi
\omega_l^{f} \eta_0)}.
\ee
 The total number of produced particles and
the total energy associated to them are given, respectively, by
\be
\left<0^p\right|N^f\left|0^p\right>=\sum_{\bf k'}\sum_{\bf k}
\left| \beta_{\bf k k'}\right|^2=\sum_{l=0}^{\infty} (l+1)^2
\frac{\mbox{sinh}^2(\pi \omega_l^{(-)} \eta_0)}{\mbox{sinh}(\pi
\omega_l^{p} \eta_0) \,\mbox{sinh}(\pi \omega_l^{f}
\eta_0)}\label{Numerototal},\ee \be E=\sum_{\bf
k}\left<0^p\right|N^f_{\bf k}\left|0^p\right>
\omega_l^{f}=\sum_{l=0}^{\infty} (l+1)^2 \frac{\omega_l^{f}
\,\mbox{sinh}^2(\pi \omega_l^{(-)} \eta_0)}{\mbox{sinh}(\pi
\omega_l^{p} \eta_0) \,\mbox{sinh}(\pi \omega_l^{f}
\eta_0)}\label{Energiatotal}.
\ee
\end{widetext}
Since we were not able to perform summation (\ref{Numerototal})
analytically, let  us make a numerical analysis of the preceding
results constructing, for instance, the graphic of the total number
of particles created $N$ {\it versus} the mass of the field
excitations $m$. Naively, we could expect that the total number of
particles created was a monotonically decreasing function of $m$.
However, as we can see from the Figure \ref{fig2}, there is a value
$m_0$ for the mass of the field at which the total number of created
particles $N(m_0)$ reaches a maximum.

It is convenient to define a width $\Delta m_0$ as it is done in a
Gaussian distribution. With this purpose, we define $m_+$ and $m_-$
such that \be N(m_0+ m_+)=N(m_0-m_-)=\mbox{e}^{-1} N(m_0), \ee so
that $ \Delta m_0:=m_+-m_-$.

\begin{figure}[!h]
\begin{center}
\newpsobject{showgrid}{psgrid}{subgriddiv=1,griddots=10,gridlabels=6pt}
\begin{pspicture}(-0.6,-0.6)(4.8,4.8)
\psset{unit=0.6}
\pspolygon[linewidth=.05](0,0)(8,0)(8,8)(0,8)
\rput{90}(-1,4){  $N$} \rput(0,-0.5){  $0$} \rput{90}(-0.5,0){ $0$}
\rput(4,-1){  $m$}
\psline[linewidth=.05](1.6666,0)(1.6666,0.25)
\psline[linewidth=.05](3.3333,0)(3.3333,0.5) \rput(3.3333,-0.5){
$100$}
\psline[linewidth=.05](5,0)(5,0.25)
\psline[linewidth=.05](6.6666,0)(6.6666,0.5) \rput(6.6666,-0.5){
$200$}
\psline[linewidth=.05](0,1.6666)(0.25,1.6666)
\psline[linewidth=.05](0,3.3333)(0.5,3.3333) \rput{90}(-0.5,3.3333){
$10\cdot 10^{4}$}
\psline[linewidth=.05](0,5)(0.25,5)
\psline[linewidth=.05](0,6.6666)(0.5,6.6666) \rput{90}(-0.5,6.6666){
$20\cdot 10^{4}$}
%
\pscurve (0., 0.) (.100, .057193) (.200, .35973) (.300, .94820)
(.400, 1.7508) (.500, 2.6586) (.600, 3.5733) (.700, 4.4153) (.800,
5.1573) (.900, 5.7657) (1., 6.2400) (1.10, 6.5910) (1.20, 6.8293)
(1.30, 6.9813) (1.40, 7.0547) (1.50, 7.0687) (1.60, 7.0307) (1.70,
6.9550) (1.80, 6.8483) (1.90, 6.7187) (2., 6.5777) (2.10, 6.4270)
(2.20, 6.2647) (2.30, 6.0923) (2.40, 5.9273) (2.50, 5.7513) (2.60,
5.5770) (2.70, 5.4110) (2.80, 5.2373) (2.90, 5.0690) (3., 4.9073)
(3.10, 4.7387) (3.20, 4.5890) (3.30, 4.4257) (3.40, 4.2770) (3.50,
4.1280) (3.60, 3.9793) (3.70, 3.8410) (3.80, 3.7013) (3.90, 3.5720)
(4., 3.4423) (4.10, 3.3162) (4.20, 3.1938) (4.30, 3.0745) (4.40,
2.9594) (4.50, 2.8469) (4.60, 2.7388) (4.70, 2.6332) (4.80, 2.5318)
(4.90, 2.4338) (5., 2.3386) (5.10, 2.2463) (5.20, 2.1581) (5.30,
2.0720) (5.40, 1.9890) (5.50, 1.9093) (5.60, 1.8321) (5.70, 1.7577)
(5.80, 1.6859) (5.90, 1.6166) (6., 1.5502) (6.10, 1.4864) (6.20,
1.4245) (6.30, 1.3649) (6.40, 1.3078) (6.50, 1.2529) (6.60, 1.1996)
(6.70, 1.1489) (6.80, 1.1001) (6.90, 1.0532) (7., 1.0080) (7.10,
.96397) (7.20, .92257) (7.30, .88300) (7.40, .84420) (7.50, .80760)
(7.60, .77247) (7.70, .73803) (7.80, .70580) (7.90, .67480) (8.,
.64443)
\pscurve[linestyle=dashed] (0., 0.) (.100, .052037) (.200, .32335)
(.300, .84200) (.400, 1.5326) (.500, 2.2929) (.600, 3.0311) (.700,
3.6830) (.800, 4.2093) (.900, 4.6123) (1., 4.8823) (1.10, 5.0380)
(1.20, 5.0930) (1.30, 5.0670) (1.40, 4.9833) (1.50, 4.8450) (1.60,
4.6813) (1.70, 4.4850) (1.80, 4.2820) (1.90, 4.0673) (2., 3.8487)
(2.10, 3.6270) (2.20, 3.4160) (2.30, 3.2078) (2.40, 3.0055) (2.50,
2.8119) (2.60, 2.6267) (2.70, 2.4508) (2.80, 2.2836) (2.90, 2.1257)
(3., 1.9768) (3.10, 1.8365) (3.20, 1.7049) (3.30, 1.5819) (3.40,
1.4663) (3.50, 1.3580) (3.60, 1.2578) (3.70, 1.1633) (3.80, 1.0755)
(3.90, .99380) (4., .91820) (4.10, .84763) (4.20, .78160) (4.30,
.72093) (4.40, .66483) (4.50, .61217) (4.60, .56383) (4.70, .51890)
(4.80, .47803) (4.90, .43940) (5., .40360) (5.10, .37137) (5.20,
.34130) (5.30, .31370) (5.40, .28803) (5.50, .26439) (5.60, .24255)
(5.70, .22255) (5.80, .20406) (5.90, .18706) (6., .17141) (6.10,
.15705) (6.20, .14384) (6.30, .13178) (6.40, .12060) (6.50, .11037)
(6.60, .10102) (6.70, .09239) (6.80, .084473) (6.90, .077240) (7.,
.070653) (7.10, .064557) (7.20, .058973) (7.30, .053817) (7.40,
.04917) (7.50, .044947) (7.60, .041037) (7.70, .037427) (7.80,
.03419) (7.90, .031204) (8., .028470)
\pscurve[linewidth=.06,linestyle=dotted, dotsep=1pt] (0., 0.) (.100,
.021064) (.200, .14187) (.300, .40090) (.400, .79457) (.500, 1.2938)
(.600, 1.8619) (.700, 2.4596) (.800, 3.0531) (.900, 3.6167) (1.,
4.1307) (1.10, 4.5800) (1.20, 4.9660) (1.30, 5.2813) (1.40, 5.5337)
(1.50, 5.7257) (1.60, 5.8647) (1.70, 5.9503) (1.80, 5.9890) (1.90,
5.9920) (2., 5.9620) (2.10, 5.9063) (2.20, 5.8367) (2.30, 5.7377)
(2.40, 5.6227) (2.50, 5.5123) (2.60, 5.3777) (2.70, 5.2433) (2.80,
5.1040) (2.90, 4.9587) (3., 4.8157) (3.10, 4.6663) (3.20, 4.5260)
(3.30, 4.3760) (3.40, 4.2377) (3.50, 4.0937) (3.60, 3.9550) (3.70,
3.8213) (3.80, 3.6830) (3.90, 3.5593) (4., 3.4320) (4.10, 3.3077)
(4.20, 3.1868) (4.30, 3.0690) (4.40, 2.9547) (4.50, 2.8435) (4.60,
2.7355) (4.70, 2.6313) (4.80, 2.5299) (4.90, 2.4320) (5., 2.3375)
(5.10, 2.2456) (5.20, 2.1573) (5.30, 2.0713) (5.40, 1.9885) (5.50,
1.9089) (5.60, 1.8317) (5.70, 1.7573) (5.80, 1.6856) (5.90, 1.6162)
(6., 1.5502) (6.10, 1.4864) (6.20, 1.4244) (6.30, 1.3651) (6.40,
1.3077) (6.50, 1.2526) (6.60, 1.1994) (6.70, 1.1491) (6.80, 1.1001)
(6.90, 1.0533) (7., 1.0081) (7.10, .96457) (7.20, .92273) (7.30,
.88247) (7.40, .84453) (7.50, .80760) (7.60, .77177) (7.70, .73843)
(7.80, .70573) (7.90, .67487) (8., .64447)
\pscurve (0., 0.) (.100, .05517) (.200, .33410) (.300, .84753)
(.400, 1.5053) (.500, 2.2017) (.600, 2.8543) (.700, 3.4143) (.800,
3.8593) (.900, 4.1927) (1., 4.4177) (1.10, 4.5607) (1.20, 4.6313)
(1.30, 4.6490) (1.40, 4.6197) (1.50, 4.5587) (1.60, 4.4723) (1.70,
4.3697) (1.80, 4.2550) (1.90, 4.1313) (2., 4.0043) (2.10, 3.8767)
(2.20, 3.7483) (2.30, 3.6213) (2.40, 3.4863) (2.50, 3.3640) (2.60,
3.2387) (2.70, 3.1161) (2.80, 2.9960) (2.90, 2.8785) (3., 2.7649)
(3.10, 2.6538) (3.20, 2.5461) (3.30, 2.4415) (3.40, 2.3403) (3.50,
2.2428) (3.60, 2.1479) (3.70, 2.0565) (3.80, 1.9684) (3.90, 1.8828)
(4., 1.8003) (4.10, 1.7216) (4.20, 1.6450) (4.30, 1.5717) (4.40,
1.5009) (4.50, 1.4329) (4.60, 1.3680) (4.70, 1.3051) (4.80, 1.2451)
(4.90, 1.1868) (5., 1.1317) (5.10, 1.0785) (5.20, 1.0278) (5.30,
.97937) (5.40, .93243) (5.50, .88767) (5.60, .84520) (5.70, .80410)
(5.80, .76547) (5.90, .72797) (6., .69210) (6.10, .65807) (6.20,
.62527) (6.30, .59473) (6.40, .56473) (6.50, .53707) (6.60, .50953)
(6.70, .48423) (6.80, .45937) (6.90, .43650) (7., .41403) (7.10,
.39340) (7.20, .37280) (7.30, .35403) (7.40, .33587) (7.50, .31855)
(7.60, .30204) (7.70, .28634) (7.80, .27142) (7.90, .25724) (8.,
.24375) \psdots  (.200, .33410) (.400, 1.5053)  (.600, 2.8543)
(.800, 3.8593)  (1.20, 4.6313)   (1.80, 4.2550)   (2.20, 3.7483)
(2.60, 3.2387)   (3., 2.7649)   (3.60, 2.1479) (4.20, 1.6450) (4.80,
1.2451) (5.40, .93243) (6., .69210) (6.60, .50953) (7.40, .33587)
\end{pspicture}
\caption{Graphic of the total number of created particles $N$ {\it
versus} the field mass $m$ for different values of parameters $A-B$,
$A+B$ and $\eta_0$. The continuous line represents the case where
$\eta_0=0.01$, $A-B=1$ and $A+B=101$. The dashed line represents the
case where $\eta_0=0.01$, $A-B=3.25$ and $A+B=101$. The dotted line
represents the case where $\eta_0=0.01$, $A-B=1$ and $A+B=51$.
Finally, the continuous dotted line represents the case where
$\eta_0=0.0115$, $A-B=1$ and $A+B=101$.} \label{fig3}
\end{center}
\end{figure}
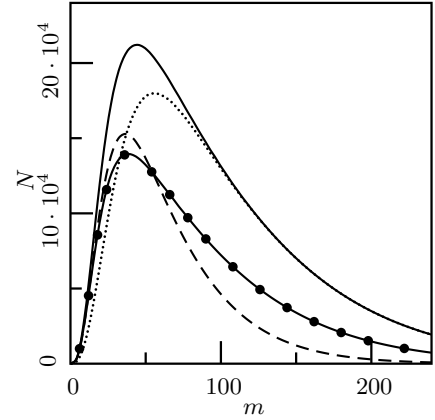

Note from equation(\ref{Numerototal}) that $N(m_0)$, $m_0$ and
$\Delta m_0$ depend only on $A-B$, $A+B$ and $\eta_0$. Despite the
fact that we were not able to get a closed analytic expression for
$m_0$, $\Delta m_0$ and $N(m_0)$ as a function of the metric
parameters $A-B$, $A+B$ and $\eta_0$, we can see numerically how
these quantities depend on those parameters. We do that by plotting
the graph of the total number of created particles as a function of
the field mass for several values of $A-B$, $A+B$ and $\eta_0$
(Figure \ref{fig3}). At first we plot a control curve (in Figure
\ref{fig3}, it is represented as a continuous line), and then we
modify one of the parameters, plot a new curve and compare to the
control one. We repeat this procedure for all parameters.

Changing the parameters $A-B$, $A+B$ and $\eta_0$ as we did in
Figure \ref{fig3} and then analyzing numerically what happens to the
graphic of $N$ {\it versus} $m$, suggests the following behaviors:
$N(m_0)$ gets larger and $\Delta m_0$ gets smaller as $A-B$ and
$\eta_0$ decrease or $A+B$ increases. On the other hand $m_0$ gets
larger as $A-B$, $A+B$ or $\eta_0$ increase.

\section{Conclusions}

In this work we computed the total number of particles created and
the total energy related to them in a $3+1$ spatially closed
Robertson-Walker space-time. By using appropriated graphics, we also
discussed  how the parameters that appear in the metric affect
the total number of particles produced. We found an unexpected
behavior of $N$ as a function of $m$, namely, starting from zero, it
increases until it reaches a maximum value at $m_0$ and after that
it decreases monotonically as $m$ increases.

The essential difference between our calculations and that presented
in {\cite{SetareRW}} is that this author does not take into account
the degeneracy factor $(l+1)^2$ that appears in equations
(\ref{Numerototal}) and (\ref{Energiatotal}). This factor must be
included due to the degeneracy in the quantum numbers $m_1$ and
$m_2$. We think our discussion may be of some help in the analysis of similar
problems, as for example, if one considers other functions $a(\eta)$ with
behaviors slightly different from that one considered here.

\section{Acknowledgments}

F. Pascoal thanks I. Waga and N. Pinto for enlightening
conversations, and FAPERJ and CNPq for financial support. C. Farina
thanks CNPq for a partial financial support.

\end{document}